\documentclass[12pt]{article}

\usepackage{amsmath}

\newcommand{\aaa}{\mathcal{A}}
\newcommand{\ajm}{\mathbf{A}}
\newcommand{\cfield}{\mathbf{C}}
\newcommand{\dff}[1]{\textsc{#1}}

\newcommand{\gnm}{\mathbf{M}}
\newcommand{\gnn}{\mathcal{N}}
\newcommand{\hh}{\mathcal{H}}
\newcommand{\ia}{\Omega}
\newcommand{\iii}{\mathcal{I}}
\newcommand{\lth}{n}
\newcommand{\rfield}{\mathbf{R}}
\newcommand{\rota}{\propto}
\newcommand{\vtr}[1]{\mathbf{#1}}

\newcommand{\erv}{\vtr{\delta}}

\begin{document}

\title{Backpropagation training in adaptive quantum networks}

\author{Christopher Altman\thanks{Applied Physics, Kavli Institute of Nanoscience, Delft University of Technology,
Postbus 5, 2600 AA Delft, The Netherlands. email: altman at
casimirinstitute.net} and
        Rom\`an R. Zapatrin\thanks{Dept. of Informatics, The State Russian
Museum, In\.{e}nernaya, 4, 191186, St.Petersburg, Russia.
              email: zapatrin at casimirinstitute.net}}


\date{}

\maketitle
\begin{abstract}
We introduce a robust, error-tolerant adaptive training algorithm
for generalized learning paradigms in high-dimensional superposed
quantum networks, or \emph{adaptive quantum networks}. The
formalized procedure applies standard backpropagation training
across a coherent ensemble of discrete topological configurations of
individual neural networks, each of which is formally merged into
appropriate linear superposition within a predefined,
decoherence-free subspace. Quantum parallelism facilitates
simultaneous training and revision of the system within this
coherent state space, resulting in accelerated convergence to a
stable network attractor under consequent iteration of the
implemented backpropagation algorithm. Parallel evolution of linear
superposed networks incorporating backpropagation training provides
quantitative, numerical indications for optimization of both
single-neuron activation functions and optimal reconfiguration of
whole-network quantum structure.

\end{abstract}

\section{Introduction}\label{intro}
Artificial neural networks are routinely applied to resolve
unstructured or multivariate machine learning problems such as
high-speed pattern recognition, image processing and associative
pattern matching tasks. Due to their novelty, however, quantum
neural networks remain relatively uncharted in the artificial
intelligence and quantum algorithms communities. Several groups
\cite{ventura,behrman,chrisley,meyer} have outlined preliminary
quantum network architectures; each novel approach contributes
significant insights towards application methodology, alternative
implementations and underlying modal interpretation -- however,
widespread and effective universal implementation of quantum neural
networks remains an open research question, as both theoretical and
experimental toolsets are still in the incipient stages of
development and maturation.

A common underlying thread shared by quantum network proposals to
date is that each implements superposition of neural transition
functions upon a single, fixed topological foundation. In this
letter, we expand upon a novel framework initially introduced in
\cite{sqnt}, which diverges from prior network models by fine-tuning
not only optimization of neural transition functions -- but by fully
reconfiguring the connective physical \emph{topology} of the quantum
network itself. The mathematical formalism employed for this
approach descends from the Rota algebraic spatialization procedure
of evolving reticular quantum structures, which was initially
developed in \cite{qstf} to address superposed topological manifolds
of spacetime foam as described in quantum gravity. Subsequently, not
only neuron weighting and \emph{transition functions} -- but also
linear superposition of the \emph{network topology} itself -- are
subject to training and revision within this coherent state space.

We formally incorporate the standard backpropagation training
algorithm initially introduced by Werbos in \cite{backprop}.
Repeated iteration of training series results in convergence of
sample output to a stable network attractor corresponding to the
lowest energy configuration space between \emph{given input} and
\emph{desired output} layer. Following convergence to this minimum,
the superposed linear network is converted upon measurement to a
conventional, classical neural network by consequent application of
the Rota algebraic projection formalism.

\section{`Linear' Neural Networks -- Dual contexts}\label{slinear}
Traditional computers execute algorithms -- that is, they follow a
specific set of instructions to arrive at a solution to a given
problem. Artificial neural networks, by contrast, learn by trial and
error: training through example. In a specific class of neural
networks -- multilayered, \emph{feedforward} neural networks --
signals are allowed to propagate only forward: there is no feedback
process in the training phase. These connection patterns are
formally classified as mathematical structures known as \dff{posets}
-- partially ordered sets, or \dff{directed acyclic graphs -- dag}s.

In this letter, we convert the primary component of artificial
neural networks -- directed acyclic graphs -- into a set of linear
matrices. The training optimization protocol is then reclassified in
terms of unitary matrices and matrix operations. The result of
training optimization is then a matrix, rather than a directed
acyclic graph. In order to recover the initial graph structure upon
derivation of the appropriate solution, the Rota algebraic
spatialization procedure \cite{fas} is applied to recover the
appropriate graph.

Herein we use the term \emph{adaptive neural networks}. However, the primary objects subject to revision and training are operators in linear spaces -- which can be realized, for instance, as quantum observables, rather than as explicitly defined neural networks. In our paper, we shall use the term `linear' under two separate contexts. The first definition implies restriction to linear artificial neural networks, as we defer application of nonlinear network optimization problems to a subsequent manuscript. In this letter, we focus solely upon optimized approach and applications of \emph{linear} neural networks.

The second usage of \emph{linearization} is the standard linear
formalism central to quantum mechanics. This is the basis of our
mathematical approach, and is outlined in further detail. The
topology of a feedforward artificial neural network, $\gnn$, is
described by the \dff{template matrix} $\ajm$ of the appropriate
directed acyclic graph, which is formed as follows:
\begin{equation}\label{etempl}
\ajm_{jk} \;=\; \left\lbrace
\begin{array}{l}
\ast,\:\mbox{if}\: j\to k \:\:\mbox{in}\:\: \gnm \cr
0,\:\mbox{otherwise}
\end{array}
\right.
\end{equation}
where $\ast$ stands for a wildcard -- any number, and the set of
such numerical matrices form an algebra \cite{rota} as it is closed
under multiplication. The main property of $\ajm$ is that the
synaptic weights `follow' it -- namely, if $\ajm_{jk}=0$, then
$w_{jk}=0$.

Returning to picture \eqref{esigprop} of signal propagation in
$\gnn$, taking various matrices $\ajm$ -- and allowing them only to
comply with the template matrix \eqref{etempl} -- we form various
products \eqref{esigprop}. The resulting set of matrices is called
the \dff{Rota algebra} of $\gnm$. It can be verified that this set
is closed under sums and matrix products, and thus qualifies as a
closed algebra. This description is explicated under greater detail
in \cite{sqnt}.

Given a feedforward neural network $\gnn$ with the set of nodes
$\gnm$, consider the linear space $\hh$, whose basis is labeled by
the elements of $\gnm$. A vector $\vec{x}\in\hh$ is associated with
a state of the network, namely the $j$-th component of $\vec{x}$ is
the activity in node $j$. Note that the state of $\gnn$ captures the
activities of \emph{all} nodes, so a convention is required to
define the default activity value of the nodes -- which are thus set to zero.

Initial activation corresponds to changing values of only a part
of nodes -- specifically those in the input layer. The signal then propagates through $\gnn$, and we describe this activation by a linear\footnote{Recall that in this paper we focus solely on linear neural networks.} operator $W$ in $\hh$. The time in which the signal requires to advance is described as the following evolution of
the state of the simulator: We start with an initial input vector
$\vtr{x}_0$, thus:
\begin{equation}\label{esigprop}
\vtr{x}_0 \;\mapsto\; \vtr{x}_1=\vtr{x}_0 W \;\mapsto\; \ldots
\;\mapsto\; \vtr{x}_{\mbox{\scriptsize out}}\;=\;
\vtr{x}_{\lth}=\vtr{x}_0 W^{\lth}
\end{equation}
following $\lth$ steps. Recall that postfix notation is employed in this
instance.

The crucial feature of the suggested approach is that signal
propagation is now described by iteration of \emph{the same}
operator $W$. From this step forward, the operator $W$, rather than
the collection of weights as in the traditional backpropagation
approach, will be subject to training.

\section{Performing standard operations in linearized form}\label{slinerr}

In this section, we reformulate classical neuron activity,
propagation, and error backpropagation methods in linear terms. A
first detail to consider is the process of signal propagation in
terms of Rota algebras. We initially have a register -- a sequence
of unique numbers defining the artificial neural network state --
and view it as a vector $\vtr{x}$ in the appropriate linear space.
Following formula \eqref{esigprop}, we consecutively act upon the
register, following elements of the Rota algebra. Initially, setup is
defined as $x=0$. Recall that the entries of $\vtr{x}$ correspond to
activities of a single neuron. When the initial signal is input, this
results in the fragment of $\vtr{x}$ corresponding to the input
layer acquiring the appropriate values, while the remainder is still
set to zero. The application of $W$ activates the neurons of the
second layer. Following this action, the first fragment bears the
input values, and the second fragment corresponds to the net inputs
of the second layer. This activation function is then applied forward
throughout the network.

The final vector, $\vtr{x}_{\lth}$, possesses the following structure,
consisting of blocks corresponding to layers -- the input block contains
corresponding input data, and the intermediate blocks contain net inputs from
the previous layers. This captures the property that as the signal propagates
through the network, the intermediate neurons retain their values. On
the other hand, this requirement is not essential --
within the same model, we may also consider artificial neural networks with simultaneous signal
inputs at the propagation point of a single input pattern. When presented with a given training set, the overall
error function may be minimized by solving the equation,
\begin{equation}\label{eexacteq}
\frac{\partial}{\partial W} AW^n-T=0
\end{equation}
One disadvantage at this point is that as the dimensionality of the
state space grows, the exact solution of this equation may become
intractable, at which point approximation methods such as gradient
descent may need to be employed. To implement this, an error
backpropagation method is required. The initial error vector $\erv$
is set to be zero. Just as in \eqref{esigprop}, the backpropagation
process can then be described. The input, $\delta_0$, is defined as
the difference between the fragment of the output $\vtr{x}_{\lth}$
of the forthpropagation process and the target vector $\vtr{t}$. The
expression for updated errors is the product of these two factors:
the first factor can be expressed as $W\vtr{x}_{\lth-t}$. The second
factor can be expressed as $\delta_t W^T$, where $W^T$ is the
transposed matrix. In accordance with agreement that the states of
$\gnn$ are diagonal matrices, rather than vectors, the product of
the two factors is well-defined. As a result,
\begin{equation}\label{ebacklin}
\erv_{t+1} \;=\; W\vtr{x}_{\lth-t}\erv_t W^T
\end{equation}
The standard weighting adjustment formula can then be written in
compact form, using the adjacency matrix $\ajm$ of the directed
acyclic graph associated with our network:
\begin{equation}\label{eupdw}
W^{\mbox{\scriptsize upd}}
\;=\;
W^{\mbox{\scriptsize ini}} + \erv_{\lth}\cdot\ajm \vtr{x}_{\lth}
\end{equation}
This procedure is then applied recurrently until required error
values converge to a minimum within predesignated boundary
tolerances. It should be noted that there is no novel formalism at
this early stage of the procedure -- this step is but a
reformulation of a standard algorithm in supervised learning.

\section{Mathematics of spatialization}\label{sspat}

The formalized spatialization procedure developed for adaptive
quantum networks is descended from the algebraic description of
quantum foam -- which describes continuously-evolving, superposed
topologies of spacetime manifold fluctuations as calculated under
quantum gravity \cite{qstf}. The key point of the spatialization
procedure is the following: consider the full matrix algebra $\aaa$
as a linear space. From this perspective, the Rota algebras $\ia$
described by \eqref{etempl} are just linear subspaces of $\aaa$.
Having a procedure which -- starting from a subset of $\aaa$ --
creates a topological space, providing the capability to discuss
superposed configurations of differing neural networks. In this
section we present a procedure which -- starting from a given
subspace of $\aaa$ -- produces a set, and endows it with a topology
that can be associated with certain acyclic directed graphs.

When the set of nodes is fixed, Rota algebras provide a suitable
machinery to describe topology changes, which are expressed in terms
of creation and annihilation of edges. In terms of template matrices
\eqref{etempl}, these operations are adding or removing asterisks
from the appropriate templates. In a fixed basis, a Rota algebra is
a subspace of a full matrix algebra $\aaa$ -- therefore adding or
removing asterisks is equivalent to adding or removing basis
vectors. Thus, any particular Rota algebra is a linear subspace of
$\ia$. The basic idea of our approach is to associate \emph{any}
subspace of $\ia$ with an appropriate directed acyclic graph.

\paragraph{Spectra -- the emergence of nodes.} The notion of a spectrum
is tightly related with that of an ideal. Suppose again that we have
an algebra of functions $\aaa$ on a set $\gnm$. Fix a point $m\in
\gnm$, and consider the subset $\iii\subseteq\aaa$ of functions $f$
such that $f(m)=0$. First, $\iii$ itself is an algebra. Furthermore,
for any $f\in\aaa$ and any $h\in\iii$ the product $f h$ is always an
element of $\iii$. Such subsets $\iii$ are called ideals of the
algebra $\aaa$. Points of $\gnm$ are in one-to-one correspondence
with the \dff{maximal ideals} of $\aaa$\footnote{A characterization
of all two-sided ideals required for spatialization was recently
provided in \cite{sorkinalg}.} and $\gnm$ is called
\dff{spectrum}\footnote{Note that there may be different spectra for
the same algebra. A simple and degenerate example is given by the
set $\cfield$ of complex numbers. Its spectrum with respect to
complex numbers consists of one point, $\gnm_\cfield=\{1\}$, while
its spectrum with respect to real numbers consists of two points,
$\gnm_\rfield=\{1,i\}$.} of $\aaa$.

In the case wherein $\aaa$ has infinite dimension, such as in a linear space, there
are different, non-equivalent and nontrivial topologies which can be
defined on $\aaa$. There are standard recipes for this step:
for example, the Zariski topology on the prime spectrum of $\ia$.
Unfortunately, on finite-dimensional algebras this topology is
always discrete -- which leaves us no chance to fit the requirement of
being non-Hausdorff. In terms of graphs, that means that the standard
recipes can be utilized to create directed arrows. Thus, we are compelled to
find another topology. For these purposes, the Rota topology was
suggested in \cite{fas}.

Suppose we are given a finite-dimensional associative -- and
non-commutative, in general -- algebra $\ia$. According to standard
conceptions and methods of modern algebraic geometry, as well as
the general algebraic approach to physics, we introduce the points
of $\ia$ as its irreducible representations (IRs). The first
step of the spatialization procedure is then creating or finding
points of $\ia$, which will become nodes of the future graph:
\begin{equation}\label{eptsirs}
\{\,\mbox{points}\,\} =
\{\,\mbox{IRs}\,\}
\end{equation}
\paragraph{Rota topology.} Denote by $\gnm$ the set of points
of $\ia$, each of which we shall associate with a prime ideal in
$\ia$. Consider two points (representations of $\ia$) $i,j\in \gnm$
and denote by $\ker{i},\ker{j}$ their kernels. Both of them, being
kernels of representations, are two-sided ideals in $\ia$, in
particular, subsets of $\ia$, hence both of the following
expressions make sense:
\[
\ker{i}\cap \ker{j} \subset \ia
\qquad\mbox{and}\qquad
\ker{i}\cdot \ker{j} \subset \ia
\]
The latter denotes the product of subsets of $\ia$: $\ker{i}\cdot
\ker{j}=\{a\in \ia\mid\, \exists u\in \ker{i},\, v\in \ker{j}:\,
uv=a\}$. Since $\ker{i},\ker{j}$ are ideals, we always have the
inclusion $\ker{i}\cdot \ker{j}\subseteq \ker{i}\cap \ker{j}$, which
may be strict or not. Define the relation $\rota$ on $\gnm$ as
follows:
\begin{equation}\label{edefrota}
i\rota j\quad
\mbox{if and only if} \quad
\ker{i}\cdot\ker{j}\neq \ker{i}\cap \ker{j}
\end{equation}
The {\dff Rota topology} is then the weakest one in which $i\rota j$
implies convergence $i\to j$ of the point $i$ to the point $j$.
Explicitly, the necessary and sufficient conditions for $i$ to
converge to $j$ in the Rota topology reads:
\begin{equation}\label{eqxy}
i\to j
\quad\mbox{if and only if}\quad
\exists k_0,\ldots,k_t,\ldots,k_n\,\mid\,
k_0=i,\,k_n=j;\,k_{t-1}\rota k_t
\end{equation}
This operation is called the transitive closure of the relation
$\rota$. Note that in general, the Rota topology can be defined upon
any set of ideals, and it is not necessary for $\ia$ to be an
algebra. Any linear subspace is suited for this purpose.

As outlined in the mathematical formalism, it
should be noted that \emph{adaptive quantum networks}, as
coherent quantum metastructures, are not standard neural networks. Rather, they are defined as superposed linear spaces, a metasystem whose subspaces consist of neural networks -- just as a superposed wavefunction possesses an identity distinct from its constituent classical
counterpart under influence of environmental interaction. In the ideal case, only following convergence to the predesignated, target network minima under backpropagation training is projection via the spatialization procedure applied to convert
the metasystem into a classical artificial neural network. As graphs are automatically produced as a consequence of this spatialization procedure, no \emph{ab initio} association of states with graphs is requisite to implement the model into a candidate physical
structure.

\section{Concluding remarks}

We have integrated standard neural network backpropagation
training \cite{backprop} into a linear, decoherence-free subspace of
superposed quantum network topologies \cite{sqnt} -- or
\emph{adaptive quantum networks} -- to simultaneously optimize both
neural transition functions and topological network configuration.
The framework is both robust and error-tolerant against local
permutations, and as such is ideally suited to applications in
rapid pattern matching, signal processing, image recognition and
associative learning.

Coherent quantum information processing architectures that are both
well-suited for implementation of adaptive quantum networks and
demonstrate promising experimental progress to date include
highly-entangled, cluster state quantum computers suggested by
Raussendorf and Briegel \cite{briegel} and implement recently by
research group headed by Zeilinger \cite{iqi}, high-dimensional
Josephson junction qubit arrays \cite{mooij}, optical lattice traps
in Bose-Einstein condensates \cite{ketterle}, thin-layer diamond
nanofilms \cite{hanson}, as well as novel materials currently under
development by research teams active worldwide.

Implementation of the training algorithm requires a relatively
limited number of physical constraints to be satisfied --
high-dimensional superposition, multiple degrees of internal
freedom, and sufficient timescale under decoherence-free coherent
state evolution. Consequently, adaptive quantum networks may be
implemented both in state-of-the-art quantum technologies, as well
as unexpectedly demonstrated within naturally-occurring dynamic
molecular complexes such as brain microtubes \cite{microtubes} and
proteomics in molecular biology \cite{fleming,cai}. Several natural
systems have been observed to exhibit interference properties and
high degrees of internal freedom -- including fullerenes
\cite{zeilinger}, tetraphenyl-porphyrin \cite{zeilinger}, small
biomolecules \cite{zeilinger}, and chlorophyll \cite{fleming}.
Contrary to conventional expections, Briegel \emph{et al.} recently
demonstrated persistent, driven entanglement in non-equilibrium
quantum systems coupled to a hot and noisy environment, in which the
presence of environmental noise played a constructive role in
maintaining recurring quantum entanglement \cite{cai}. As such,
promising candidates in coherent quantum information processing
technologies, condensed-matter systems, and biophysics are all
potential areas of investigation for experimental implementation of
adaptive quantum networks \cite{vedral,bouwmeester}.

Superposed quantum network topologies -- \emph{adaptive quantum
networks} -- as outlined in this paper, are unique in allowing for
simultaneous training of both transition functions and optimization
of whole-network topological structure. Quantitative backpropagation
training provides predictive indications for optimal reconfiguration
of network performance, which can be applied towards a broad class
of problems currently addressed by classical neural networks.
Continued progress will focus upon predictive simulation methods to
determine convergence series in standardized sample classes, complex
and chaotic attractor states under dynamic entanglement coupling
strengths, alternative training paradigms, and nonlinear dynamics in
the presence of environmental coupling through partially-induced
weak measurements.

\paragraph{Acknowledgements.} CTA wishes to thank Adele Engele Behar for her
continued inspiration, encouragement and support. Lev Levitin and
Jack Tuszynski provided excellent discussions and feedback in the
early stages of the paper. Both CTA and RRZ acknowledge the
hospitality of the organizers, and particularly Jaros\l aw Pykacz.
RRZ was supported under auspices of research grant RFFI 07-06-00119.



\end{document}